\documentclass{PoS}
\usepackage{subfigure}

\title{GASP - Galway Astronomical Stokes Polarimeter}

\ShortTitle{GASP- Galway Astronomical Stokes Polarimeter }

\author{Patrick P. Collins\\
        Centre for Astronomy,School of Physics, National University of Ireland, Galway, Ireland\\
        E-mail: \email{p.collins1@nuigalway.ie}}

\author{Brendan Shehan\\
        Centre for Astronomy,School of Physics, National University of Ireland, Galway, Ireland\\
        E-mail: \email{bshehan-nuig@yahoo.com}}

\author{Michael Redfern\\
        Centre for Astronomy,School of Physics, National University of Ireland, Galway, Ireland\\
        E-mail: \email{mike.redfern@nuigalway.ie}}

\author{Andrew Shearer\\
        Centre for Astronomy,School of Physics, National University of Ireland, Galway, Ireland\\
        E-mail: \email{andy.shearer@nuigalway.ie}}

\abstract{The Galway Astronomical Stokes Polarimeter (GASP) is an ultra-high-speed, full Stokes, astronomical imaging polarimeter based upon a Division of Amplitude Polarimeter. It has been developed to resolve extremely rapid stochastic ($\sim$ms) variations in objects such as optical pulsars, RRATs and magnetic cataclysmic variables. GASP has no moving parts or modulated components, so the complete Stokes vector can be measured from just one exposure - making it unique to astronomy. Furthermore the time required for the determination of the full Stokes vector is limited only by the time resolution of the detectors used and the incident photon fluxes. GASP utilizes a modified Fresnel rhomb, which acts as a highly achromatic quarter wave plate and a beamsplitter (referred to as an RBS). Here we present a description of how the DOAP works, some of  the optical designs for the polarimeter, and give some preliminary results. Calibration is an important, and difficult issue with all polarimeters, but particularly in astronomical polarimeters. We give a description of  calibration techniques appropriate to this type of polarimeter, particularly the Eigenvalue Calibration Method of Compain \& Drevillon}

\FullConference{Polarimetry days in Rome: Crab status, theory and prospects\\
		 October 16-17 2008\\
		 Rome, Italy}

\begin{document}
	
\section{Introduction}
\label{intro}

Optical polarimetry, traditionally relies upon elements such as rotating half-wave plates and Wollaston prisms to determine the polarisation. Such a methodology works for objects which exhibit polarisation that is static, slowly changing or has well defined periodic behavior.   Normal emission from optical pulsars clearly falls into this category, but stochastic phenomena - such as the optical counterpart to giant radio emission \cite{shear03} or rotating radio transient (RRAT) emission \cite{mc06}  - present a difficult problem for polarimetry.  Photo-polarimeters, among the fastest polarimeters, can obtain Stokes I, Q, and U in $\sim \mu$s but this may still not be fast enough for stochastic phenomna. GASP is designed to address this problem by measuring simultaneously all Stokes vectors and thereby enabling a determination of both linear and circular polarisation in a single exposure.  Similar considerations can be given to studies of other stochastic phenomena such as flickering associated with CVs. Most importantly, GASP will be in position for target of opportunity investigations such as polarisation measurements of the afterglow of gamma ray bursts and searches for the optical counterparts of new gamma-ray pulsars which will be discovered by the Fermi GRO. In this paper we describe the design criteria for the polarimeter.

\section{Division of Amplitude Polarimetry (DOAP)}
\label{sec:DOAP2}
DOAP (see Azzam \cite{azzam82}) can measure the entire Stokes vector instantaneously. This is achieved by splitting the incoming light across a specially coated beamsplitter to divide the light to linear and circular components, then onto a polarizing beam splitter, normally a Wollaston prism. As a result, four beams of different intensities will relate linearly to the input Stokes vector as follows, $\mathbf{I=AS}$ where $\mathbf{I}$ is a $4\times1$ intensity vector, $\mathbf{A}$ is a $4\times4$ matrix for the system and $\mathbf{S}$ is the input vector retrieved from $\mathbf{S=IA^{-1}}$. Only the photon fluxes and time-resolution of the detector used will limit the temporal resolution of the Stokes vector obtained. 

\begin{figure}[h]
\centering
\subfigure[Original DOAP from Azzam \cite{azzam82}]{
\includegraphics[width=7cm]{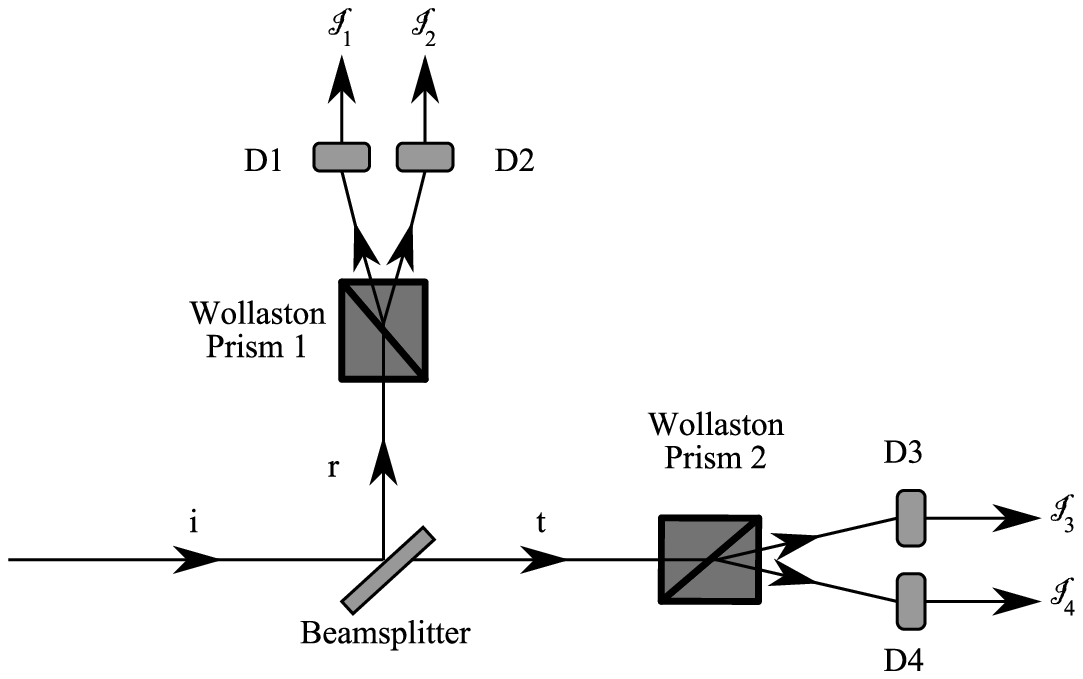}
\label{fig:DOAP}
}
\subfigure[An early polarimeter design following developments of the Compain\cite{compain98} polarimeter]{
\centering
\includegraphics[width=3cm]{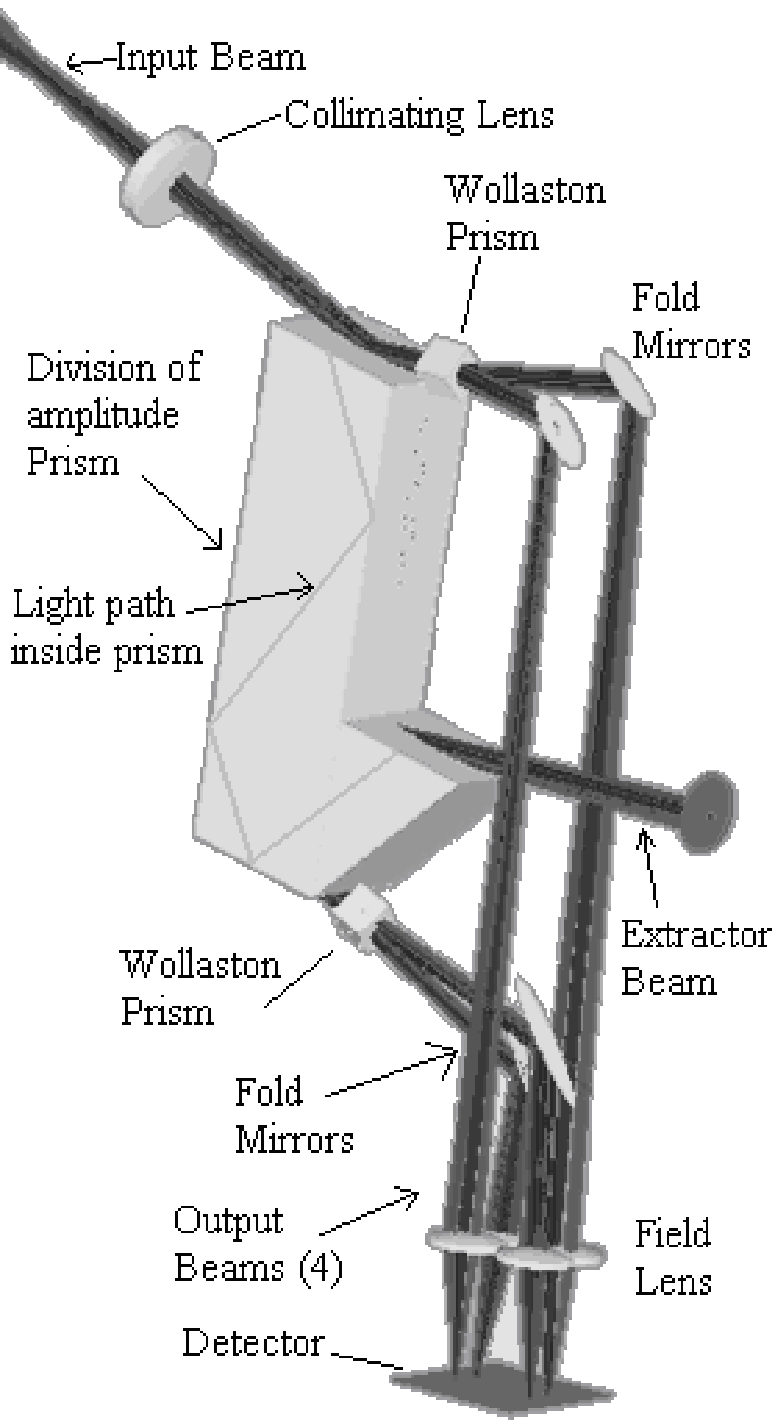}
\label{fig:gasp_oldbw}
}
\label{DOAPS}
\caption{One of the  optical designs where the Wollaston splits the light immediately after the RBS and each field has its own optics onto the imager.}
\end{figure}

\begin{figure}[h]
	\centering
		\includegraphics[width=12cm]{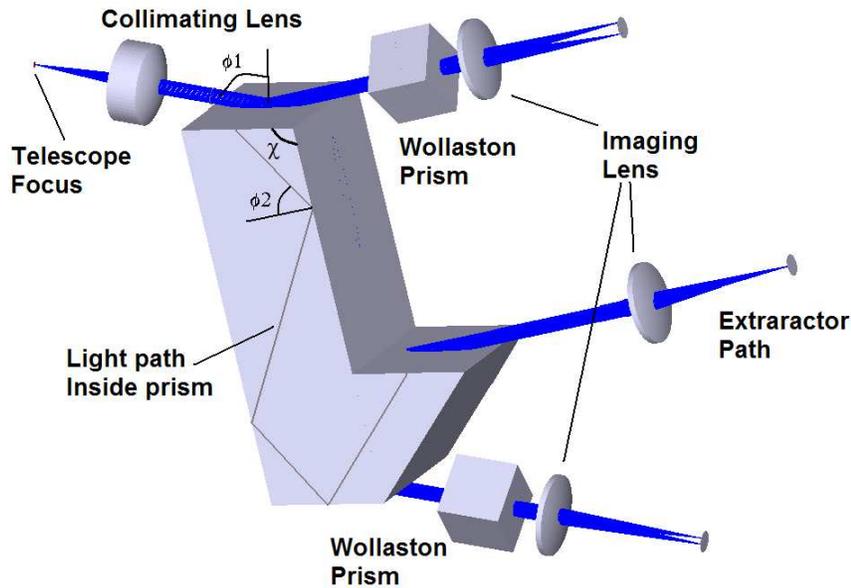}
	\caption{Schematic optical layout for the GASP }
	\label{fig:optical_layout}
\end{figure}

Compain \& Drevillon \cite{compain98} describe a DOAP in which a modified Fresnel rhomb, similar to that in Fig \ref{fig:gasp_oldbw} and \ref{fig:optical_layout}, is used. We have based the  Galway Astronomical Stokes Polarimeter (GASP) upon this design. The prism is designed to act as a highly achromatic quarter wave plate (QWP) and a retarding beamsplitter (RBS). Particular attention was paid to the design of this prism as the optomisation of the system matrix is best when the intensity of both reflected and transmitted beams are equal. Utilizing Snells law, and the prism geometry and the equation for calculating the retardance induced by internal reflection (see Born and Wolf \cite{bw}) the prism angle ($\chi$) can be solved. The prism angle ($\chi$) is designed to yield a total quarter wave redardance for two internal reflections($\phi_{2}$).  At the same time the angle of incidence, $\phi_1$, is such that both reflected and transmitted beams are of equal intensity. When each beam is passed through a Wollaston prism these beams are split into two further polarisation states. The resulting four beams are re-imaged  enabling their intensities to be extracted and consequently we can determine the full polarimetric characterisation of the input beam. 

\section{Optical Design}
\label{sec:opt_design}
Fig \ref{fig:optical_layout}. Shows the schematic design diagram for GASP. The incoming beam is split three ways with two used for polarimetry and the third, the extracted waste light, for guidance and finding. A  Wollaston prism is used to split the two polarimetric channels  and each individual beam is picked off to either four individual APD modules or two EMCCD detectors. The overall quantum efficiency (DQE) of each beam is 8-12\% with the higher number being for the EMCCD system and the lower for the APDs.

\section{Calibration}
\label{sec:Calib}
Polarimeters are among the hardest instruments to calibrate in astronomy. Normally the  instrumental polarization is determined and is subtracted from the data. This means that the polarimeter must be designed to have the analyser as soon as possible in the light path as other elements may depolarize the input. GASP works by purposely polarizing the light before the analysers so a different calibration method is used.  
A Polarization State Generator (PSG) Generates four linearly independent, optimized \cite{azzam88,compain99}, Stokes vectors which are	 measured in the lab. The PSG vectors are passed through the system and four respective intensity vectors are recorded. The system matrix, $\mathbf{A}$, is calculated by $\mathbf{A=IW^{-1}}$. Where $\mathbf{W}$ and $\mathbf{I}$ are $4\times4$ matrices formed by the 4 PSG and intensity vectors respectively. No instrumental polarization will exist as this is a total polarimeter and this has been calibrated out. The calibration is verified by measuring polarization and zero polarization standard stars.

\subsection{ECM Calibtation}
\label{sec:ECM}
Calibration of the polarimeter is achieved through the Eigenvalue Calibration Method (ECM) \cite{compain99}. All one has to do is measure the Mueller matrices of 4 known high quality samples, Air, a polarizer at 0$^\circ$ and 90$^\circ$ and a QWP at 28$^\circ$. Eigenvalues analysis of the 4 Mueller matrices a linear mapping algorithm, unambiguously determines the error correction matrices for $\mathbf{A}$ and $\mathbf{W}$. The ECM method achieves 1\% accuracy, consequently the polarimeter can then measure polarization with similar accuracy. 

\section{Current Progress}
\label{sec:progress}
\begin{figure}[h]
	\centering
		\includegraphics[width=10cm]{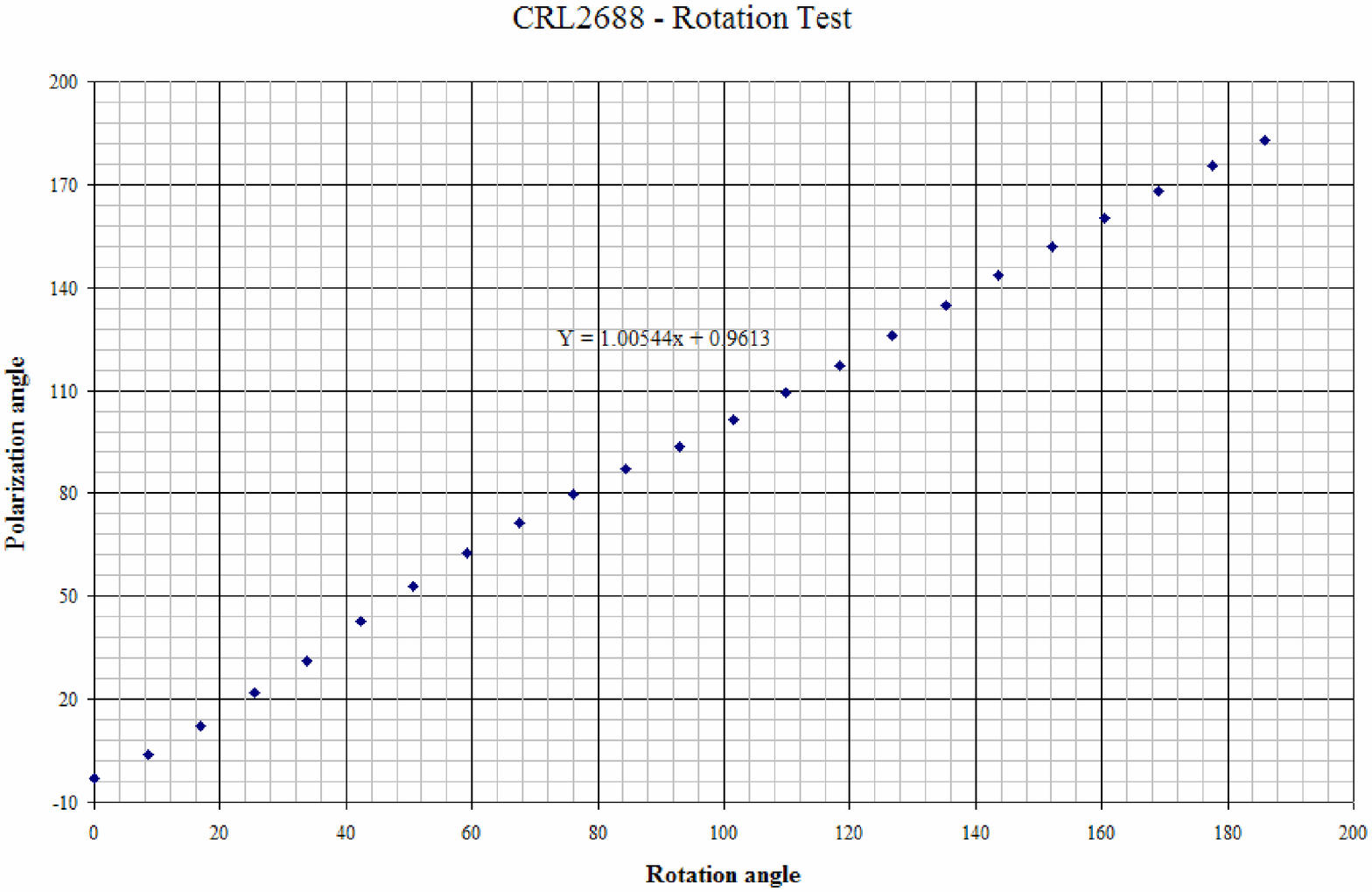}
	\caption{Polarization angle versus field rotation for the rotation test of a polarised source.}
	\label{f1}
\end{figure}

\begin{figure}[h]
	\centering
		\includegraphics[width=10cm]{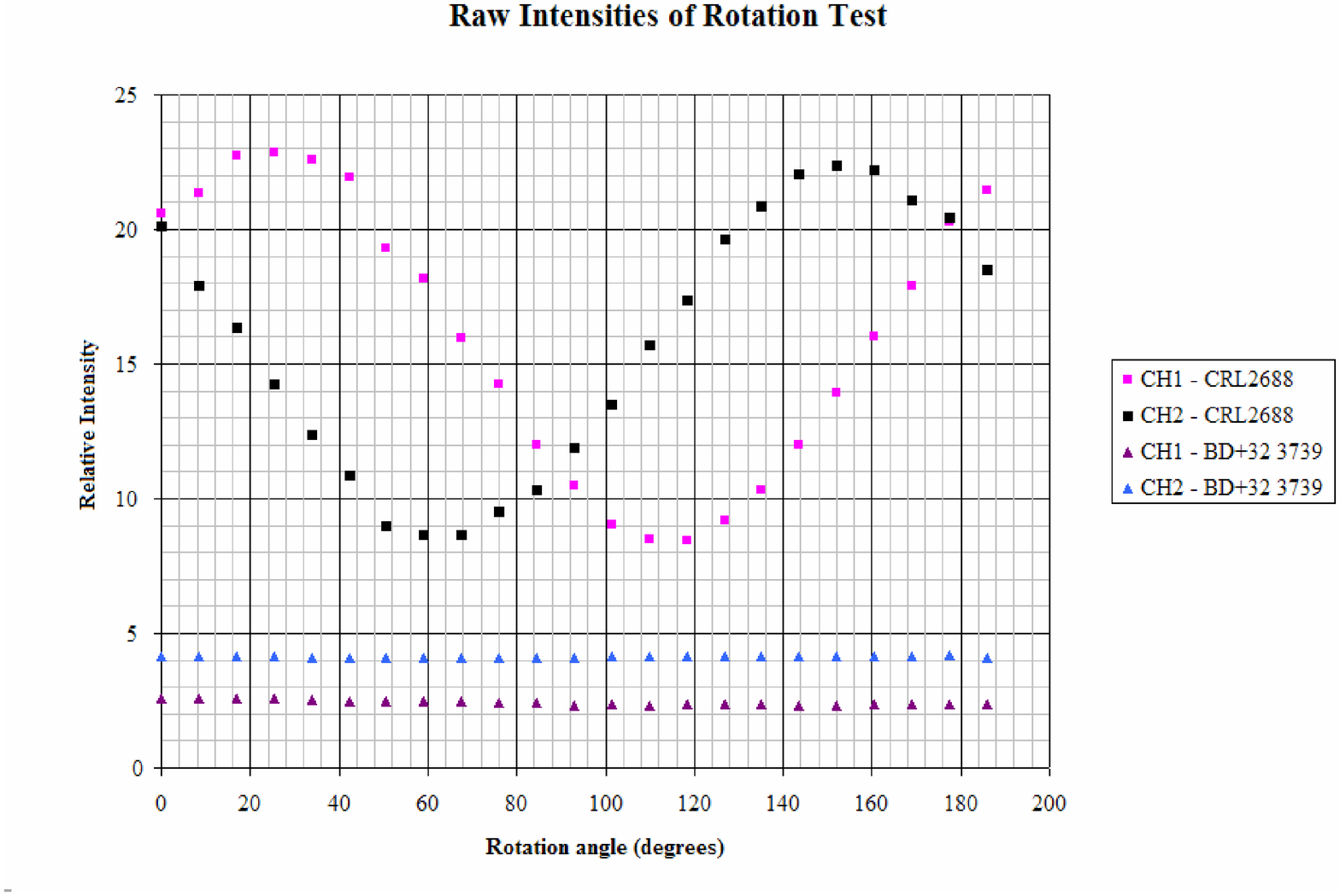}
	\caption{Raw channel outputs (ch1 and ch2) for the rotation test of a polarised (CRL 2688) and  unpolarised (BD+32 3739) sources.}
	\label{f2}
\end{figure}

GASP has been tested on the 1.5m Cassini telescope at Loiano Observatory, Bologna. Here we show the results on two standards - one a 50\% polarised source CRL 2688 and the other an unpolarised standard - BD+32 3739. Fig \ref{f1} shows an instrument rotation test where we show the measured angle of polarisation as a function of instrument orientation with respect to parallactic angle on the sky. Fig \ref{f2} shows the the measured intensities of both sources again as a function of relative instrument orientation. The significance of this test is that under rotation the degree of linear polarisation should remain constant whilst the angle of polarisation should vary precisely with the rotation angle, even though each of the four output beams will vary sinusoidally. This is demonstrated in upper two panels, thus proving that the A matrix can be correctly calibrated to allow accurate inversion of the $\mathbf{I}$ vector, yielding the $\mathbf{S}$ vector for individual frames. The fluctuations in Stokes I and U/Q were consistent with our model for the propagation of Poisson noise. Fig {f3} shows the determined degree of polarisation for a series of 0.1 s exposures of CRL 2688 using the Cassini telescope. This will be equivalent to a $\approx$ 5 ms exposures of the Crab pulsar on a 8m class telescope.

\begin{figure}[h]
	\centering
		\includegraphics[width=10cm]{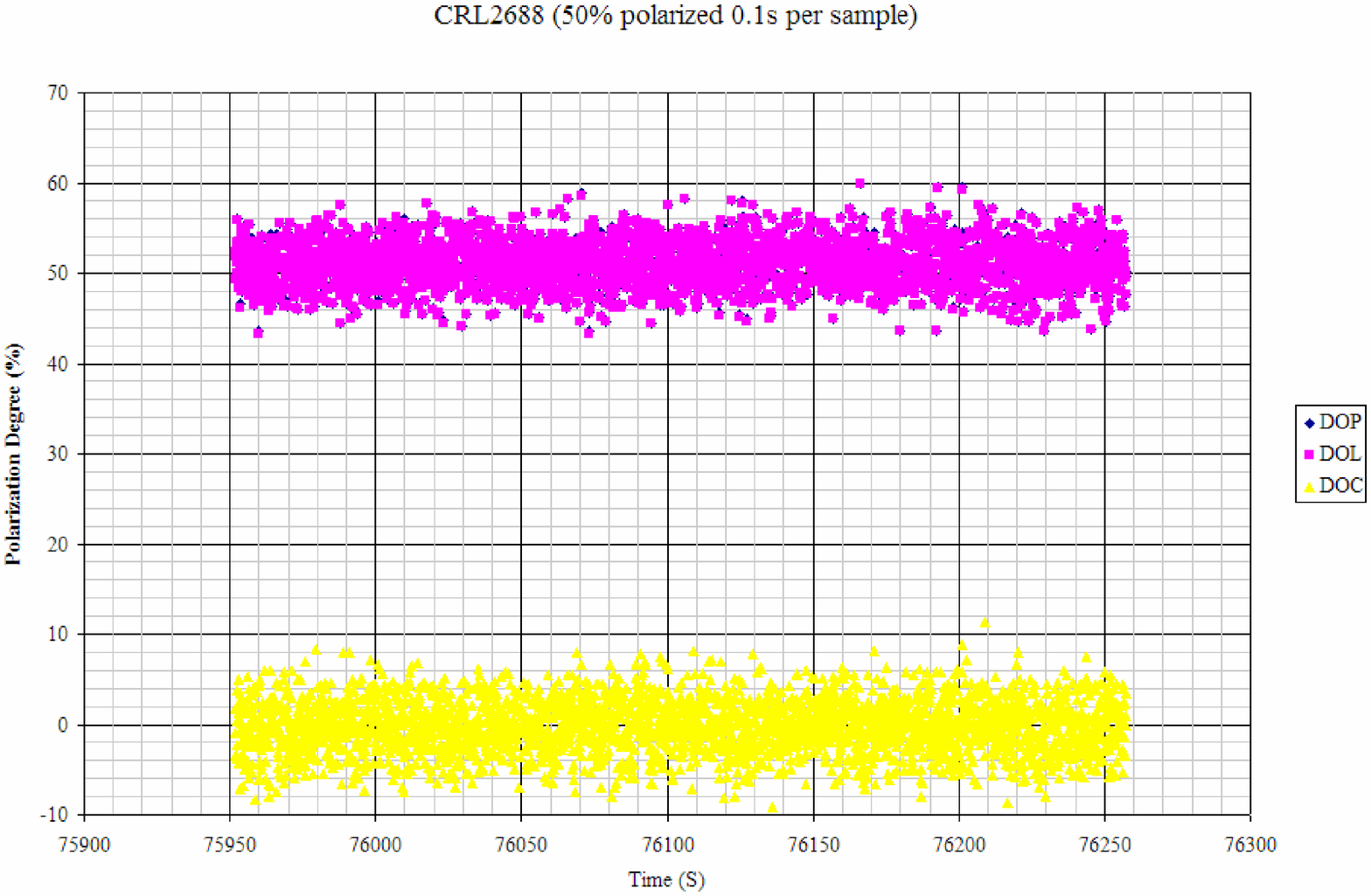}
	\caption{Degree of polarization versus time for CRL2688 showing the noise characteristics of the series with 0.1 s exposures}
	\label{f3}
\end{figure}

\section{Acknowledgements}
\label{sec:Acknowl}
Alexander Goncharov, Nicholas Devaney, Chris Dainty, David Lara, from NUI, Galway's Applied Optics group are thanked for their contribution and assistance. PC thanks SFI and the NUI Galway Fellowship programme for their financial assistance. Science Foundation Ireland, who supported the development of GASP under grant 05/RFP/PHY0045

Antonello de Martino, for advice and support and for the loan of equipment including the original ``Compain \& Drevillon'' RBS



\bibliography{refC}{}
\bibliographystyle{abbrv}


%

\end{document}